\documentclass[twocolumn]{aastex62}
\pdfoutput=1 
\usepackage{amsmath,amstext,gensymb}
\usepackage[T1]{fontenc}
\usepackage[figure,figure*]{hypcap}
\graphicspath{{./}{figures/}}

\submitjournal{AAS}

\shortauthors{Wang et al.}

\begin{document}

\title{TOI-4495: A Pair of Aligned, Near-Resonant Sub-Neptunes that Likely Experienced Overstable Migration}

\author[0000-0003-3015-6455]{Mu-Tian Wang}
\affiliation{Institute for Astronomy, University of Hawai`i, 2680 Woodlawn Drive, Honolulu, HI 96822, USA}
\affiliation{School of Astronomy and Space Science, Nanjing University, Nanjing 210023, China.}
\affiliation{Key Laboratory of Modern Astronomy and Astrophysics, Ministry of Education, Nanjing, 210023, People’s Republic of China}

\author[0000-0002-8958-0683]{Fei Dai}
\affiliation{Institute for Astronomy, University of Hawai`i, 2680 Woodlawn Drive, Honolulu, HI 96822, USA}

\author[0000-0001-5162-1753]{Hui-Gen Liu}
\affiliation{School of Astronomy and Space Science, Nanjing University, Nanjing 210023, China.}
\affiliation{Key Laboratory of Modern Astronomy and Astrophysics, Ministry of Education, Nanjing, 210023, People’s Republic of China}

\author[0000-0003-1298-9699]{Kento Masuda}
\affiliation{Department of Earth and Space Science, Osaka University, Osaka 560-0043, Japan}

\author[0000-0001-8638-0320]{Andrew W. Howard}
\affiliation{Department of Astronomy, California Institute of Technology, Pasadena, CA 91125, USA}

\author[0000-0003-1312-9391]{Samuel Halverson}
\affiliation{Jet Propulsion Laboratory, California Institute of Technology, 4800 Oak Grove Drive, Pasadena, CA 91109, USA}

\author[0000-0002-0531-1073]{Howard Isaacson}
\affiliation{{Department of Astronomy,  University of California Berkeley, Berkeley CA 94720, USA}}
\affiliation{Centre for Astrophysics, University of Southern Queensland, Toowoomba, QLD, Australia}

\author[0009-0004-0455-2424]{Elina Y. Zhang}
\affiliation{Institute for Astronomy, University of Hawai`i, 2680 Woodlawn Drive, Honolulu, HI 96822, USA}

\author[0000-0003-3868-3663]{Max Goldberg}
\affiliation{Laboratoire Lagrange, UMR7293, Universit\'e C\^ote d'Azur, CNRS, Observatoire de la 
C\^ote d'Azur, Bouldervard de l'Observatoire, 06304, Nice Cedex 4, France}

\author[0000-0003-3860-6297]{Huan-Yu Teng}
\affiliation{Korea Astronomy and Space Science Institute
776 Daedeok-daero, Yuseong-gu, Daejeon 34055, Republic of Korea}
\affiliation{Institute for Astronomy, University of Hawai`i, 2680 Woodlawn Drive, Honolulu, HI 96822, USA}
\affiliation{CAS Key Laboratory of Optical Astronomy, National Astronomical Observatories, Chinese Academy of Sciences, Beijing 100101, China}

\author[0000-0003-3856-3143]{Ryan A. Rubenzahl}
\affiliation{Center for Computational Astrophysics, Flatiron Institute, 162 Fifth Avenue, New York, NY 10010, USA}

\author[0000-0003-3504-5316]{Benjamin Fulton}
\affiliation{NASA Exoplanet Science Institute/Caltech-IPAC, MC 314-6, 1200 E California Blvd, Pasadena, CA 91125, USA}

\author[0000-0003-0967-2893]{Erik A. Petigura}
\affiliation{Department of Physics \& Astronomy, University of California Los Angeles, Los Angeles, CA 90095, USA}

\author[0000-0002-8965-3969]{Steven Giacalone}
\affiliation{Department of Astronomy, California Institute of Technology, Pasadena, CA 91125, USA}

\author[0000-0002-9305-5101]{Luke Handley}
\affiliation{Department of Astronomy, California Institute of Technology, Pasadena, CA 91125, USA}

\author[0000-0001-9911-7388]{David W. Latham}
\affiliation{Center for Astrophysics \textbar Harvard \& Smithsonian, 60 Garden St, Cambridge, MA 02138, USA}

\author[0000-0001-6637-5401]{Allyson Bieryla}
\affiliation{Center for Astrophysics \textbar Harvard \& Smithsonian, 60 Garden St, Cambridge, MA 02138, USA}

\author[0000-0002-6525-7013]{Ashley Baker}
\affil{Caltech Optical Observatories, Pasadena, CA, 91125, USA}

\author[0009-0002-2419-8819]{Jerry Edelstein}
\affil{Space Sciences Laboratory, University of California Berkeley, Berkeley, CA 94720, USA}

\author[0009-0004-4454-6053]{Steven R. Gibson}
\affil{Caltech Optical Observatories, Pasadena, CA, 91125, USA}

\author{Kodi Rider}
\affil{Space Sciences Laboratory, University of California Berkeley, Berkeley, CA 94720, USA}

\author[0000-0001-8127-5775]{Arpita Roy}
\affiliation{Astrophysics \& Space Institute, Schmidt Sciences, New York, NY 10011, USA}

\author{Chris Smith}
\affil{Space Sciences Laboratory, University of California Berkeley, Berkeley, CA 94720, USA}

\author[0000-0002-6092-8295]{Josh Walawender}
\affiliation{W. M. Keck Observatory, 65-1120 Mamalahoa Hwy, Waimea, HI 96743}

\author[0000-0003-2196-6675]{David Rapetti}
\affiliation{USRA, Washington DC, 20024 USA/NASA Ames Research Center, Moffett Field, CA 94035 USA}

\author[0000-0002-4715-9460]{Jon M. Jenkins}
\affiliation{NASA Ames Research Center, Moffett Field, CA 94035, USA}
\affiliation{Research Institute for Advanced Computer Science, Universities Space Research Association, Washington, DC 20024, USA}



\author[0000-0002-4265-047X]{Joshua N. Winn}
\affiliation{Department of Astrophysical Sciences, Princeton University, 4 Ivy Lane, Princeton, NJ 08544, USA}



\begin{abstract}
\noindent We report the discovery of a sub-Neptune and a Neptune-like planet (R$_b=2.48^{+0.14}_{-0.10}$ R$_\oplus$, R$_c=4.03^{+0.23}_{-0.15}$ R$_\oplus$) orbiting the F-type star TOI-4495. 
The planets have orbital periods of 2.567 days and 5.185 days, lying close to a 2:1 mean-motion resonance (MMR). Our photodynamical analysis of the TESS light curves constrains the planetary masses to M$_b=7.7\pm1.4$ M$_\oplus$ and M$_c=23.2\pm4.7$ M$_\oplus$. 
The measured masses and radii indicate the presence of volatile-rich gaseous envelopes on both planets.
The Rossiter–McLaughlin effect and the Doppler shadow of TOI-4495 c reveal a well-aligned orbit with a projected stellar obliquity of $\lambda=-2.3^{+8.3}_{-7.8}$\degree. Combined with the low mutual inclination constrained by the photodynamical analysis ($\Delta I<8.7$\degree), the planetary orbits are likely coplanar and aligned with the host star's spin axis. 
We show that the planets are near, but not in, the 2:1 MMR, with a circulating resonant angle. We also find substantial free eccentricity for the inner planet, TOI-4495 b ($e_b=0.078^{+0.02}_{-0.013}$). Given the observed proximity to the 2:1 resonance and the more massive outer planet, TOI-4495 b and c are particularly susceptible to resonant overstability, which in turn can explain the observed eccentricity by converting resonantly excited eccentricity into free eccentricity. However, additional mechanisms (e.g., planetesimal scattering) may be required to further excite the eccentricity by $\sim4\%$. To prevent tidal damping from reducing the eccentricity below the observed level over the star’s lifetime (1.9 Gyr), the reduced tidal quality factor of TOI-4495 b must be $Q' \gtrsim10^5$, consistent with the presence of a thick envelope on the planet.

\end{abstract}

\keywords{Exoplanet astronomy (486), Transit timing variation method (1710), Orbital resonances (1181)}

\section{Introduction}
Near-resonant planetary systems provide valuable insights into planetary formation and dynamical evolution. Their orbital architectures point to a history of convergent migration within protoplanetary disks, followed by stabilization through mean-motion resonances (MMRs) \citep{Terquem_2007,Cresswell,IdaLin2008}. Observationally, resonant systems are more prevalent among the youngest stars and tend to be systematically younger than non-resonant systems \citep{Dai2024,Schmidt2024}. Moreover, near-resonant planets often exhibit lower bulk densities than non-resonant counterparts of similar radii, suggesting either formation beyond the snow line with subsequent inward migration, or a history of avoiding disruptive giant impacts during the final stages of assembly \citep{Mills2017_ttvmass,Leleu2024,Chen2024}.

These findings support a scenario in which resonant configurations established during the disk phase are frequently disrupted over time by dynamical processes \citep[e.g.,][]{Izidoro,Izidoro2021,Li2025}. Most observed near-resonant planet pairs display period ratios that deviate from exact commensurability by a few percent \citep{Fabrycky2014} and exhibit small but non-zero free eccentricities that is not related to resonant interactions \citep{Lithwick_ttv,Wu2013}. Such characteristics suggest that even near-resonant systems, despite preserving quasi-pristine orbital architectures, are commonly reshaped by subsequent perturbations. Possible mechanisms include disk turbulence \citep{Batygin2015_capture,Adams2008,Goldberg2022,Chen2025_turbulent}, disk-edge effects \citep{Liu2017,Huang_Ormel,Hansen2024,Liu2022}, planet–planet interactions \citep{Choksi_TTVphase}, and scattering with residual planetesimals \citep{Chatterjee_2015,Wu2024}. This complex interplay among disk-driven migration, post-disk evolution, and planetary composition motivates detailed investigations of individual near-resonant systems to unravel their formation histories.

In this work, we present TOI-4495, a system consisting of a sub-Neptune and a Neptune-like planet near the 2:1 MMR, orbiting an F-type star on close-in orbits ($P_b=2.56$ day, $P_c=5.18$ day). The near-resonant configuration produces transit timing variations (TTVs) for both planets in the TESS observations, which we analyze through detailed dynamical modeling. We find significant eccentricity that is not induced by resonant interactions. We also conducted follow-up spectroscopic observations of the outer planet, TOI-4495 c, revealing that its projected orbital axis is aligned with the stellar spin axis ($\lambda=-2.3^{+8.3}_{-7.8}~^\circ$). In addition, we report a non-detection of excess H$\alpha$ absorption during transit. With the dynamical model constrained by TTVs, we demonstrate that the planets lie near, but not within, the 2:1 MMR. Finally, we show that the dynamical state cannot be explained by disk migration alone, suggesting that additional dynamical excitation processes may be at play.

The paper is organized as follows. Section \ref{sec:obs} describes the observations used in this work. Section \ref{sec:stellar_param} presents the characterization of the stellar properties. Section \ref{sec:RM} contains the analysis of the Rossiter–McLaughlin (RM) effect and the H$\alpha$ transmission observations of TOI-4495 c. Section \ref{sec:ttv} presents the analysis of the transit light curves, TTVs, and the dynamical modeling. We summarize our findings in Section \ref{sec:result} and conclude in Section \ref{sec:concl}.

\section{Observation \label{sec:obs}}
\subsection{TESS}
TOI-4495 (TIC 120826158) was observed by the TESS mission \citep{Ricker} in Sectors 14, 40, 41, 53, 54, 80, and 81 between UT 2019 July 18 and 2024 July 25. Our analysis is based on the 2-min cadence light curves reduced by the TESS Science Processing Operations Center \citep[SPOC;][]{jenkinsSPOC2016}, available from the Mikulski Archive for Space Telescopes. We tested both the Simple Aperture Photometry \citep[SAP;][]{Twicken2010,Morris2020} and the Presearch Data Conditioning Simple Aperture Photometry \citep[PDCSAP;][]{Smith2012,Stumpe2012,Stumpe2014} versions of the light curves, and present results based on the PDCSAP data. To minimize the influence of anomalous data, we excluded cadences with nonzero quality flags.

\subsection{TRES}
We obtained two high-SNR spectra of TOI-4495 with the Tillinghast Reflector Echelle Spectrograph \citep[TRES;][]{fureszTRES} in October 2021. TRES, located at the Fred L. Whipple Observatory on Mt. Hopkins in southern Arizona, is a fiber-fed echelle spectrograph covering wavelengths from 390 to 910 nm with a resolving power of approximately 44,000. The spectra were reduced following the procedure outlined in \cite{buchhave2010}.

\subsection{KPF}
The Keck Planet Finder \citep[KPF;][]{Gibson2024} is a high-resolution, high-stability echelle spectrograph on the 10 m Keck I telescope at the W. M. Keck Observatory. It operates over two channels spanning wavelengths from 445 to 870 nm, with a median resolving power of 98,000. Wavelength calibration is achieved using ThAr and UNe lamps together with a Menlo Systems laser frequency comb \citep{Gibson2024}. On the night of UT 2023 July 8, we observed a total of 86 spectra of TOI-4495 during a transit of TOI-4495 c. We achieved a typical signal-to-noise ratio (SNR) of 240 at 5500 \AA{} and 270 at 7500 \AA{}. The spectra were reduced with the KPF Data Reduction Pipeline\footnote{https://github.com/Keck-DataReductionPipelines/KPF-Pipeline}.

\section{Stellar Parameter \label{sec:stellar_param}}
To determine the spectroscopic parameters of TOI-4495 from the TRES spectra, we used the Stellar Parameter Classification tool \citep[SPC;][]{buchhave2012}. SPC derives stellar properties by cross-correlating the observed spectra with a grid of synthetic spectra generated from Kurucz atmospheric models \citep{kurucz1992}. The resulting effective temperature $T_{\rm eff}$, surface gravity $\log g$, iron abundance [Fe/H], and $v\sin i$ are listed in Table \ref{tab:stellar_para}.

We combined the Gaia parallax measurement \citep{gaiaDR3} with our spectroscopic parameters to refine the stellar properties. The parallax of $7.5884 \pm 0.0117$ mas provides an independent constraint on the stellar radius via the Stefan–Boltzmann law. Using the spectroscopic effective temperature, the $K$-band magnitude, and the Gaia parallax, we computed the stellar radius with the \texttt{Isoclassify} package \citep{Huber17}, which performs isochronal fitting by combining spectroscopic and astrometric constraints. The fit employed MESA Isochrones \& Stellar Tracks \citep[MIST;][]{Choi}. The posterior distributions of the stellar parameters are summarized in Table~\ref{tab:stellar_para}.

We estimated the stellar age using both isochrone fitting and gyrochronology. Spectral energy distribution (SED) modeling with \texttt{EXOFASTv2} \citep{Eastman2013} yields an isochronal age of $1.8^{+0.8}_{-0.7}$ Gyr. From the TESS light curve, we measured a stellar rotation period of $10.2 \pm 2.2$ days using the auto-correlation function \citep{McQuillan2014}. Applying the gyrochronology relations of \citet{Bouma_2023} gives an age estimate of $1.9^{+0.5}_{-0.7}$ Gyr. The two methods yield consistent results within uncertainties.

\begin{deluxetable*}{lcc}[htbp]
\tablecaption{Stellar Parameters of TOI-4495} 
\tablewidth{1.0\textwidth}
\label{tab:stellar_para}
\tablehead{
\colhead{Parameters} & \colhead{Median and 68.3\% Credible} Interval & \colhead{Reference}}
\startdata
{\bf Stellar Parameter}\\
TIC ID                                                  & 120826158                     & A \\
R.A.                                                    & 19:05:27                      & A \\
Dec.                                                    & +37:01:33.56                  & A \\
$V$ (mag)                                               & 9.531 $\pm$ 0.003             & A \\
$K$ (mag)                                               & 8.217 $\pm$ 0.018             & A \\
Effective Temperature $T_{\text{eff}} ~(K)$             & $6210\pm70$                   & B \\
Surface Gravity $\log~g~(\text{cm~s}^{-2})$             &$4.29 \pm 0.040$               & B \\
Iron Abundance $[\text{Fe/H}]~(\text{dex})$             &$0.18\pm 0.10$                 & B \\
Rotational Broadening $v~\text{sin}~i_*$ ~(km~s$^{-1}$) &$6.8\pm 0.6$                   & B \\
Stellar Mass $M_{\star} ~(M_{\odot})$                   &$1.247\pm0.045$                & B \\
Stellar Radius $R_{\star} ~(R_{\odot})$                 &$1.309\pm0.059$                & B \\
Stellar Density $\rho_\star$ (g cm$^{-3}$)              &$0.77\pm0.10$                  & B \\
Rotation Period $P_{\rm rot}$ (days)                    &10.2$\pm 2.2$                  & B \\
Isochronal Age (Gyr)                                    &$1.8^{+0.8}_{-0.7}$            & B \\
Gyrochronology Age (Gyr)                                &$1.9^{+0.5}_{-0.7}$            & B \\
Parallax $\pi$ (mas)                                    &$7.5884\pm0.0117$              & C \\
\hline
\enddata
\tablecomments{A:TICv8 \citep{Stassun}; B: this work; C: \citet{gaiaDR3}}
\end{deluxetable*}

\section{Rossiter-Mclaughlin Effect and H$\alpha$ transmission \label{sec:RM}}

\subsection{Rossiter-McLaughlin Effect}

\begin{figure*}[!htb]
\center
\includegraphics[width = 1.\textwidth]{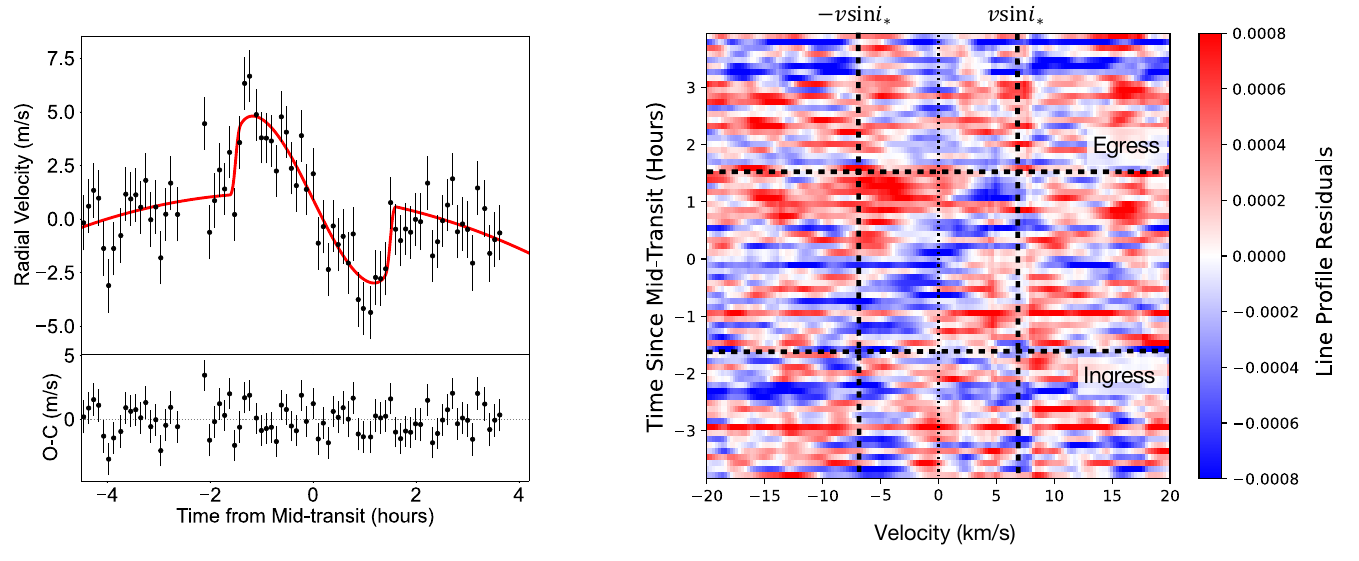}
\caption{\emph{Left:} radial velocity variations during transit of TOI-4495 c (Rossiter-Maclaughlin effect). The black points are KPF data, and the red curve shows the best-fit RM model with sky-projected stellar obliquity of $\lambda=-2.3^{+8.3}_{-7.8}$\degree. The bottom panel shows the residuals of data subtracted by the best-fit model. \emph{Right:} line-profile residuals as a function of time and velocity. The vertical black dashed lines mark the $v\sin i_*=6.8\pm0.6$ km/s of the host star. The horizontal black dashed lines indicate the ingress and egress timing of transit. The blue diagonal feature, extending from -6.8 km/s at ingress to 6.8 km/s at egress, represents the Doppler shadow of TOI-4495\,c and is consistent with a well-aligned orbit.
}
\label{fig:RM_DT}
\end{figure*}

The Rossiter–McLaughlin (RM) effect \citep{Rossiter1924,Mclaughlin1924} is a spectroscopic anomaly during planetary transits that reveals the sky-projected stellar obliquity. As the planet successively blocks the blueshifted and redshifted hemispheres of the rotating star, the integrated stellar spectrum shows a time-varying anomalous Doppler shift across the transit.
With the KPF observations taken during the transit of TOI-4495 c, we measured the projected stellar obliquity as follows. We used our best-fit transit model from the TESS light curves to assist in modeling the RM effect. Specifically, we modeled the phase-folded, TTV-adjusted TESS transits of planet c simultaneously with the RM effect. The RM model included the time of conjunction as a free parameter to account for the large TTVs. Reassuringly, the best-fit mid-transit time from the RM measurement confirmed the TTV of planet c and followed the expected trend from the TESS data (Figure \ref{fig:ttv}). Our RM model closely followed the prescription of \cite{Hirano2011}. In addition to the standard transit parameters modeled in Section \ref{sec:transit_modeling}, the RM model also required the following parameters: the sky-projected obliquity $\lambda$, the projected rotational velocity, a linear function of time to describe the local radial velocity (RV) trend with an offset $\gamma$, and the local linear and quadratic RV variations $\dot{\gamma},~\ddot{\gamma}$. An RV jitter term was also included to capture any additional astrophysical or instrumental noise. No clear evidence of correlated (red) noise was seen in the RM residuals; we therefore adopted a simple $\chi^2$ likelihood function with a penalty term for the jitter parameter \citep[e.g.,][]{Howard2013}. The best-fit model was obtained using the Levenberg–Marquardt method implemented in the Python package \texttt{lmfit} \citep{LM}.

To sample the posterior distribution, we used the Markov Chain Monte Carlo (MCMC) technique implemented in \texttt{emcee} \citep{emcee}. We initialized 128 walkers near the best-fit model and ran them for 10,000 steps, corresponding to more than 50 times the autocorrelation length for all parameters. The RM measurement and best-fit model is shown in the left panel of Figure \ref{fig:RM_DT}. We found that TOI-4495 c has a sky-projected obliquity of $\lambda=-2.3^{+8.3}_{-7.8}$\degree, consistent with zero. Combining $v\sin i_*$, the stellar radius, and the stellar rotation period from TESS, we constrained the stellar inclination to be $\cos i_*>0.67$ at 95\% credible lower limit \citep{Masuda_vsini}. Following the procedure of \cite{Albrecht2021}, we found that the true obliquity is consistent with zero, with an upper limit of 39$^\circ$ at the 95\% credible level.

The Doppler shadow \citep[e.g., ][]{Cameron2010_dopplershadow} is another way to measure the sky-projected stellar obliquity. This refers to the localized distortion in the stellar line profile caused by a transiting planet blocking part of the rotating stellar surface. As the planet moves across the star, this distortion traces a moving “shadow” in velocity space.
We perform Doppler shadow analysis on TOI-4495 c using the method of \cite{Dai1726}. We subtracted the average out-of-transit line profile from the in-transit line profiles. The right panel of Figure \ref{fig:RM_DT} shows the line-profile residuals as a function of time and stellar-centric velocity. The Doppler shadow of TOI-4495 c appears as a blue diagonal feature in the center of the diagram. Given the lower SNR of the Doppler shadow signal, we adopt the stellar obliquity constraint from the RM analysis.

\subsection{Atmospheric Erosion Traced by H$\alpha$ Absorption}

High-energy radiation from a host star can erode a planet’s atmosphere through mechanisms such as photoevaporation, in which stellar X-ray and extreme ultraviolet photons heat the upper atmospheric layers, driving bulk atmospheric escape \citep[e.g.,][]{OwenWu2017}. This effect is strongest for planets on close-in orbits around young, magnetically active stars \citep{Ribas2005}. Transmission spectroscopy at high spectral resolution provides a powerful tool to detect atmospheric escape in exoplanets \citep[e.g.,][]{Spake2018,Zhang2023}. If a planet possesses an extended, escaping atmosphere, additional absorption can be observed during transit. This excess absorption can be disentangled from variations in the stellar line profiles, and the signal undergoes Doppler shifts during transit due to the planet’s orbital motion across the stellar disk.

We extracted the transmission spectrum of TOI-4495\,c around the H$\alpha$ line (6564.75\,\text{\AA}). A total of 43 spectra were obtained, spanning approximately 2.1 hours before and after mid-transit. We removed a telluric line near $\sim 6566.6$\,\AA{} by fitting a Gaussian profile to each exposure after continuum subtraction; in any case, this telluric feature is well separated from the H$\alpha$ line. 
To search for evidence of atmospheric escape, we computed the equivalent width of the H$\alpha$ line within a 2\,\AA{} bandpass as a function of time. No significant time-variable excess absorption was detected, suggesting the absence of, or strongly suppressed, atmospheric escape. Given the mature age of the system (1.9 Gyr), this null detection is consistent with theoretical expectations that photoevaporation occurs predominantly early in a planet’s evolutionary history \citep{Ribas2005,Owen,France2025}.


\section{Transit Timing Variations\label{sec:ttv}}

\subsection{Extracting Transit Timing\label{sec:transit_modeling}}

\begin{figure*}
\center
\includegraphics[width = .49\textwidth]{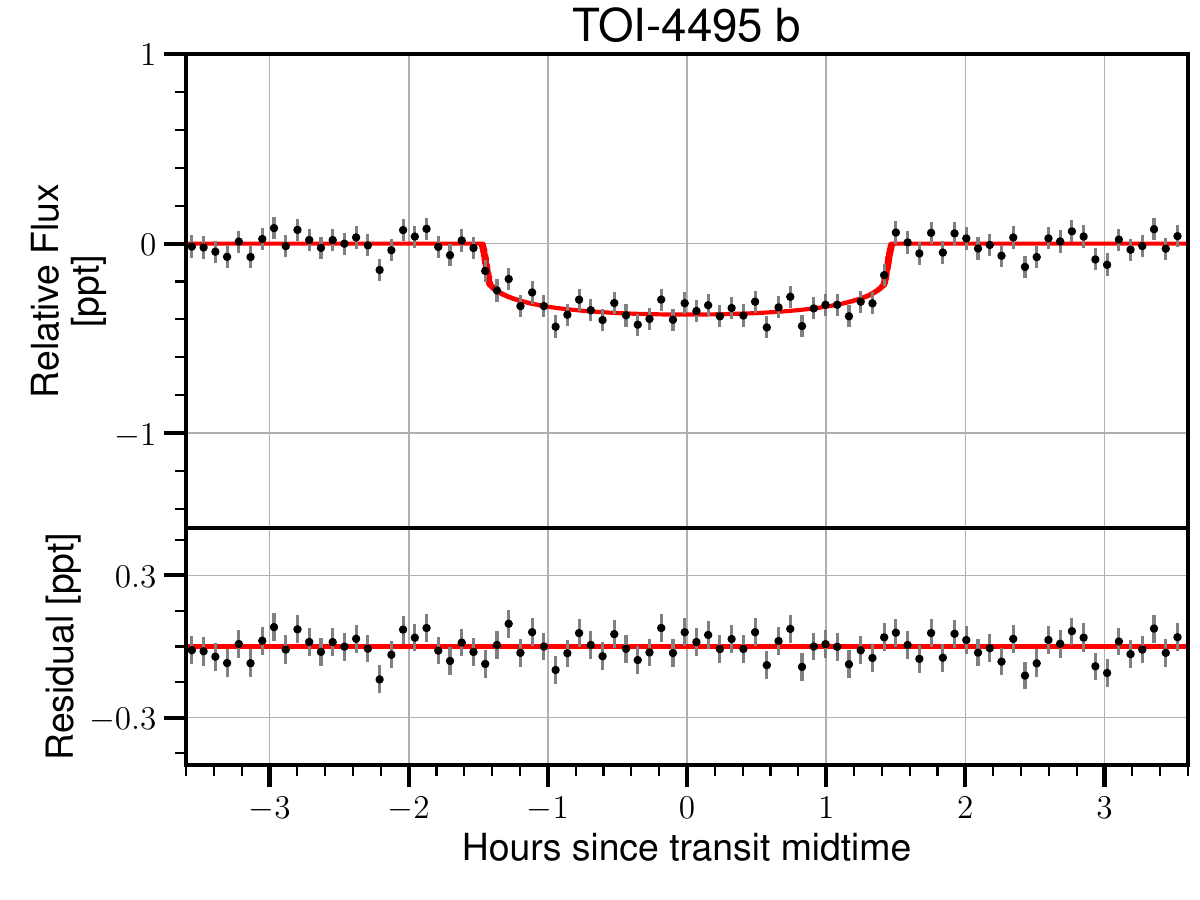}
\includegraphics[width = .49\textwidth]{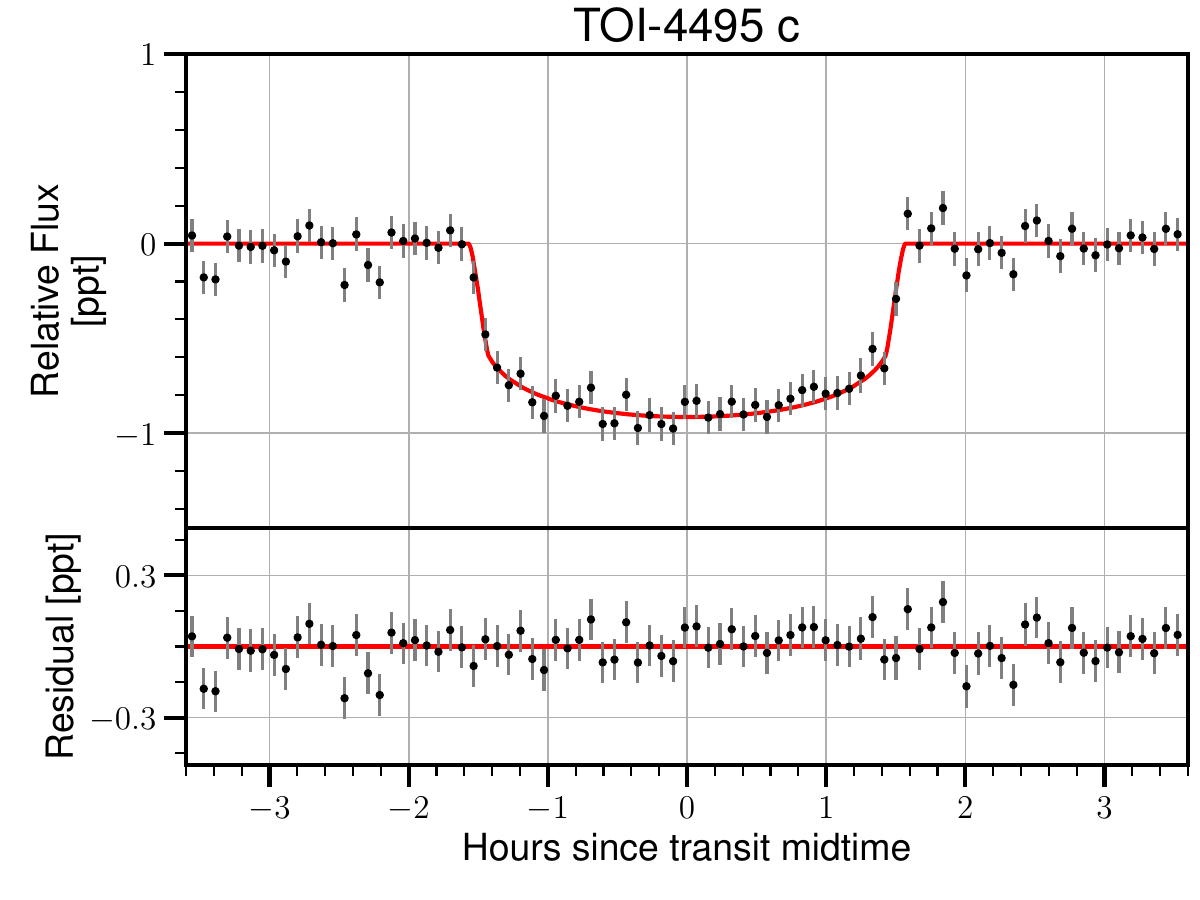}
\caption{Phase folded {\it TESS} transit light curves of TOI-4495 b and c after accounting for TTVs, binned to 5 min. Red line is the best-fit photodynamics model (Section \ref{sec:photodynamics})}
\label{fig:light_curve}
\end{figure*}

We modeled the transit light curves using the Python package \texttt{batman} \citep{Kreidberg2015}, adopting the precise stellar density derived in Section~\ref{sec:stellar_param} as a prior. A quadratic limb-darkening law was implemented using the $q_1$ and $q_2$ parameterization of \citet{Kipping} to enable efficient sampling. Gaussian priors with a width of 0.3 were placed on the limb-darkening coefficients, centered on the theoretical values computed with \texttt{EXOFAST} \citep{Eastman2013}. The mean stellar density and the two limb-darkening coefficients were treated as global parameters shared across all planets in the system.

For each planet, the fitted transit parameters included the orbital period $P_{\rm orb}$, time of conjunction $T_c$, planet-to-star radius ratio $R_p/R_*$, scaled semi-major axis $a/R_*$, impact parameter $b$, the orbital eccentricity $e$, and the argument of the pericenter $\omega$.

Before fitting the transits, we removed stellar variability and instrumental systematics from the TESS light curves by fitting a cubic spline with a 0.5-day knot spacing. Data points within twice the transit duration ($T_{14}$) of each event were excluded from the spline fit, with transit timing variations (TTVs) accounted for in subsequent iterations. The original light curve was then divided by the spline model. Visual inspection confirmed that the detrending preserved the transit shapes without introducing distortions.

We first fitted each planet’s transit assuming a constant orbital period, obtaining a best-fit model using the Levenberg–Marquardt algorithm implemented in \texttt{lmfit} \citep{LM}. This model served as a template for fitting the mid-transit times of individual events, where only the mid-transit time and three coefficients of a quadratic baseline were allowed to vary. In Sectors 80 and 81, overlapping transits of the two planets were handled by fitting them simultaneously. The total flux loss was modeled as the sum of the individual losses, without accounting for possible planet–planet eclipses \citep[e.g.,][]{Hirano_planet_planet}.

Accurate detrending of the light curve and isolation of transit windows rely critically on precise estimates of both TTVs and transit durations. After obtaining an initial dynamical fit, we refined the transit windows using the more accurate TTV model to generate the final light curves for the photodynamical analysis in Section~\ref{sec:photodynamics}. For each transit, we masked the in-transit data and fitted the out-of-transit flux within a window spanning $3\times T_{14}$ using polynomial functions up to fourth order. The adopted baseline is the lowest–order polynomial function whose Bayesian Information Criterion \citep[BIC;][]{Schwarz1978} is lower than that of any higher–order polynomial model.

\subsection{Evidence for Free Eccentricity \label{sec:free_ecc}}

\begin{figure*}
    \centering
    \includegraphics[width=1.0\linewidth]{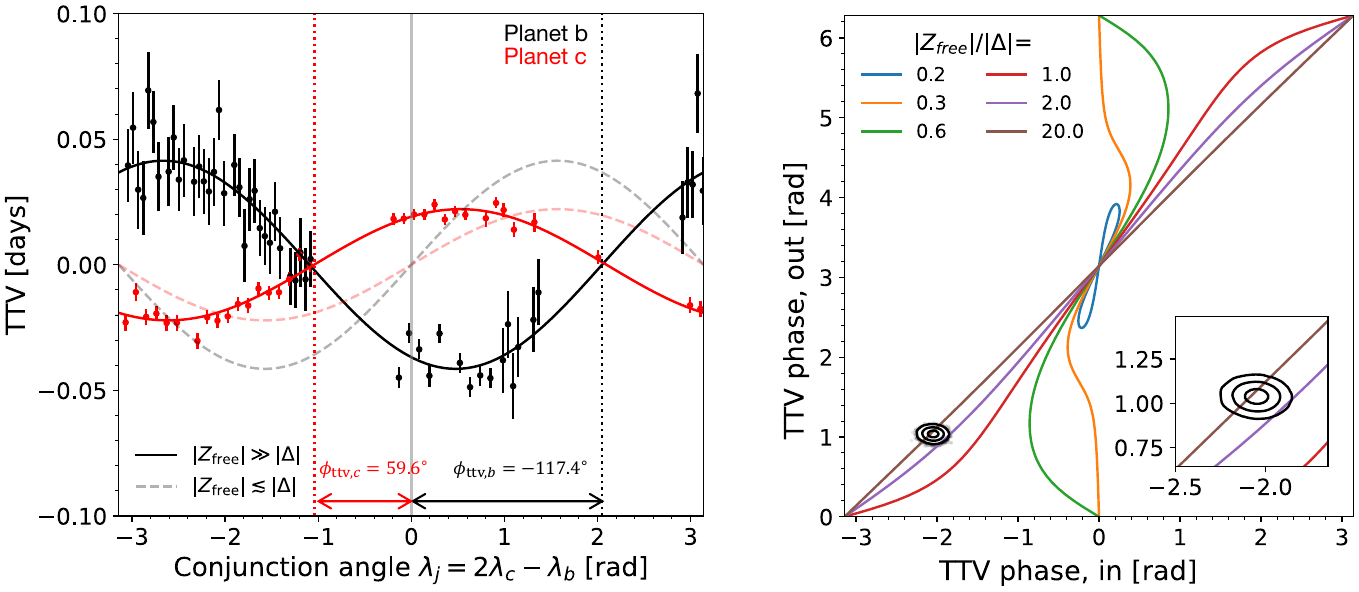}
    \caption{Left: TTVs of planet b (black) and planet c (red) phased to the conjunction angle $\lambda_j = 2\lambda_c - \lambda_b$. If the free eccentricity $Z_{\rm free}$ is small compared to $\Delta$, no TTV phase shift is expected (dashed lines; TTV = 0 when $\lambda_j = 0$), which is inconsistent with the observations (solid lines). We measure TTV phases (vertical dotted lines) of $\phi_{b} = -117.4 \pm 3.5^\circ$ and $\phi_{c} = 59.5 \pm 2.2^\circ$ for planets b and c, respectively. Right: TTV phases of two planets near the 2:1 MMR. Colored curves show theoretical TTV phases for different values of $Z_{\rm free}$. When $Z_{\rm free}$ is comparable to or larger than $\Delta$, non-zero TTV phases arise. The TTV phase posteriors are shown as contours corresponding to the 1, 2, and 3$\sigma$ credible levels. For TOI-4495, the TTV data favor $|Z_{\rm free}|$ values larger than $\Delta$ by a factor of a few. See Section~\ref{sec:free_ecc} for further discussion.
}
    \label{fig:ttv_phaseshift}
\end{figure*}

The eccentricities of planets near an MMR can be decomposed into a forced component, driven by perturbations from the companion, and a free component that reflects the intrinsic orbital non-circularity \citep{Lithwick_ttv}. The free eccentricity is encoded in the TTV phases \citep{Lithwick_ttv,Choksi_TTVphase}, and our analysis presented below finds TOI-4495 planets exhibit significant free eccentricity.

Planets near mean-motion resonances exhibit TTV patterns that approximate sinusoidal functions, which take the form: 
\begin{align}
    t_{1,2} &= nP_{1,2}+T_{1,2} \\ \nonumber
    &+ {\rm Real(}\mathcal{V}_{1,2})\sin\lambda_j+{\rm Imag(}\mathcal{V}_{1,2})\cos\lambda_j ,
\end{align}
where subscripts 1 and 2 refer to the inner and outer planets. $P$ and $T$ are the mean orbital period  and reference midtime of transit. For a first-order MMR, $\lambda_j=j\lambda_2-(j-1)\lambda_1$ is the conjunction angle for planets near a $j:(j-1)$ resonance, and $\lambda_{1,2}$ are the mean longitudes of the planets. Therefore the TTV signals will have the periodicity at super-period $P_{\rm sup}=(j/P_2-(j-1)/P_1)^{-1}$. The complex quantity $\mathcal{V}_{1,2}$ has the form
\begin{align}
    \mathcal{V}_1 &= \frac{P_1}{\pi}\frac{\mu_2}{j\Delta}\alpha^{-1/2}\left(-f-\frac{j-1}{j}\alpha^{-3/2}\frac{3Z^*_{\rm free}}{2\Delta}\right) \\ 
    \mathcal{V}_2 &= \frac{P_2}{\pi}\frac{\mu_1}{j\Delta}\left(-g+\frac{3Z^*_{\rm free}}{2\Delta}\right),
\end{align}
where $\Delta=(j-1)P_2/(jP_1)-1$ is the pair’s distance from nominal resonance. $f=-1.19+2.2\Delta$ and $g=0.428-3.69\Delta$ are Laplacian coefficients near the 2:1 MMR with first-order corrections in $\Delta$. The complex quantity $Z^*_{\rm free}$, which controls the TTVs, is the conjugate of the combined free eccentricity, defined as
\begin{equation}
\label{eq:comb_ecc}
    Z_{\rm free}=fz_{\rm 1,free}+gz_{\rm 2,free},
\end{equation}
where $z_{1,2}=e_{1,2}\exp(i\omega_{1,2})$ are the complex eccentricities, and $e$ and $\omega$ are the eccentricity and longitude of pericenter of each planet. The free eccentricity can be isolated by subtracting the resonantly forced component $z_{\rm forced}$ from the total eccentricity,
\begin{align}
    z_{\rm 1, forced} &= \frac{1}{j\Delta}\mu_2f\left(\frac{P_1}{P_2}\right)^{1/3} \exp(i\lambda_j), \label{eq:forced_ecc_1} \\ 
    z_{\rm 2, forced} &= \frac{1}{j\Delta}\mu_1g \exp(i\lambda_j). \label{eq:forced_ecc_2}
\end{align}

The TTV phase is defined as the argument of the complex quantity, $\phi_{1,2}={\rm arg}(\mathcal{V}_{1,2})$. If the free eccentricity is small ($|Z_{\rm free}|\ll\Delta$), the TTVs follow a sinusoidal function with minimal phase shift (i.e., ${\rm TTV}=0$ when $\lambda_j=0$) \citep{Lithwick_ttv,Deck2015}. This behavior is not observed in the left panel of Figure \ref{fig:ttv_phaseshift}. Instead, the non-zero free eccentricity induces a measurable phase shift, which we constrained to be $\phi_{b}=-117.4\pm3.5^\circ$ and $\phi_{c}=59.5\pm2.2^\circ$, following the approach of \cite{Lithwick_ttv}. The right panel of Figure \ref{fig:ttv_phaseshift} illustrates the expected ranges of TTV phase for the inner and outer planets as a function of $|Z_{\rm free}|/\Delta$ near the 2:1 MMR. Small values of $Z_{\rm free}$ result in phases concentrated near $(0,\pi)$. Significant deviations from $(0,\pi)$ occur only when $Z_{\rm free}\gtrsim\Delta$, in which case both planets’ TTV phases shift toward the diagonal line in phase space. The precisely measured TTV phases therefore require a nonzero $Z_{\rm free}$.

As shown in the right panel of Figure~\ref{fig:ttv_phaseshift}, the measured TTV phases are consistent with any $Z_{\rm free}$ larger than $2\Delta$, meaning that the first-order analytic treatment alone does not provide an upper limit on $Z_{\rm free}$. The upper limit can instead be constrained by including the contribution from the second-order MMR (4:2) term in the analytic TTV model. This term has a period equal to half the super-period and an amplitude smaller than the first-order term by a factor of $Z_{\rm free}$ \citep{Deck2015,Hadden2017}. In the residuals of the left panel of Figure \ref{fig:ttv_phaseshift}, we do not see evidence for this second-harmonic trend, implying that $Z_{\rm free}$ is bounded by the noise amplitude. A more rigorous constraint on $Z_{\rm free}$ will be obtained through full $N$-body modeling, as described in the following section.


\subsection{Photodynamics}
\label{sec:photodynamics}

We performed photodynamical modeling with \texttt{jnkepler}\footnote{https://github.com/kemasuda/jnkepler}, which computes the relative flux loss at observed time for both transiting planets while accounting for their mutual gravitational interactions. The system is initialized using osculating orbital parameters of  planet-to-star mass ratio, orbital period $P$, eccentricity $e$, argument of pericenter $\omega$, nodal angle $\Omega$, and time of inferior conjunction $T_{\rm conj}$. With a symplectic integrator, the model solves the planetary motions and determines the positions and velocities at each transit center. The in-transit motion of each planet was assumed to be linear. The flux loss $F$ was calculated using the solution vector formalism of \citet{Agol2020}, with further implementation details described in \citet{Masuda2024}. 
 
The likelihood function $\mathcal{L}$ has the form of 
\begin{equation}
\mathcal{L} = \mathcal{N}(f; \mu,\Sigma)
\end{equation}
The $f$ is the residual of observed flux substracted by the light curve model $F$. $\mathcal{N}(x; \mu,\Sigma)$ is a multivariate normal distribution with mean $\mu$ and coveriance matrix $\Sigma$. We modeled stellar variability with a Matérn-3/2 Gaussian process kernel, therefore the $ij$ elements of $\Sigma$ is:
\begin{align}
    \Sigma_{ij} &= \sigma_{\rm TESS}^2 \, \delta_{ij} \nonumber \\ 
    &+ a^2 \left( 1 + \frac{\sqrt{3}\,|t_i - t_j|}{c} \right)
    \exp\!\left(-\frac{\sqrt{3}\,|t_i - t_j|}{c}\right)
\end{align}
$\delta_{ij}$ is the Kronecker delta function, $t_{i,j}$ is the observing epoch of $i$-th or $j$-th datapoint, $a,c$ are the characteristic amplitude and length scale of photometric variability, and $\sigma_{\rm TESS}$ is a white-noise parameter representing the effective combination of photometric uncertainties and additional jitter of TESS light curves.
 The log-likelihood was evaluated using the \texttt{JAX} \citep{Bradbury2018} interface of \texttt{celerite2} \citep{celerite2}. 

In the fit, we used the normalized and detrended flux obtained in Section \ref{sec:transit_modeling}. For the 20-second cadence data, we bin to 2 min. Transit observations are insensitive to the absolute nodal angles of individual planets and only constrain the relative nodal angle \citep{Ragozzine2010}. Therefore, we fixed $\Omega_c = 0$, and restricted $|\Omega_b| < 1$ rad. We also required both planets to have $b>0$. Stellar masses $M_*$ and radius $R_*$ were given Gaussian priors determined from spectroscopy measurements in Table \ref{tab:stellar_para}. The rest of parameters were given uniform priors, which are summarized in Table \ref{tab:fitted_and_derived_physical_parameter}.

\texttt{jnkepler} is implemented in \texttt{JAX} and supports gradient-based inference method such as Hamiltonian Monte Carlo \citep[HMC;][]{Duane1987,Betancourt2017}, which is more effective in exploring the long, thin, and curving degeneracies in posterior parameter space that are hard to sample. Therefore we carry out the posterior sampling with the No-U-Turn Sampler implemented in \texttt{NumPyro} \citep{Phan2019}. We ran eight chains for 1,000 tuning steps followed by 1,000 sampling steps, yielding a total of 8,000 posterior samples. Convergence diagnostics showed Gelman–Rubin statistics \citep{BB13945229} of $\hat{R} < 1.01$ and an effective sample size of at least 360. The median values and 68\% credible intervals for the fitted and derived parameters are reported in Table~\ref{tab:fitted_and_derived_physical_parameter}.


\section{Result\label{sec:result}}

\begin{figure*}
    \hspace{-0.5cm}
    \includegraphics[width = 2.\columnwidth]{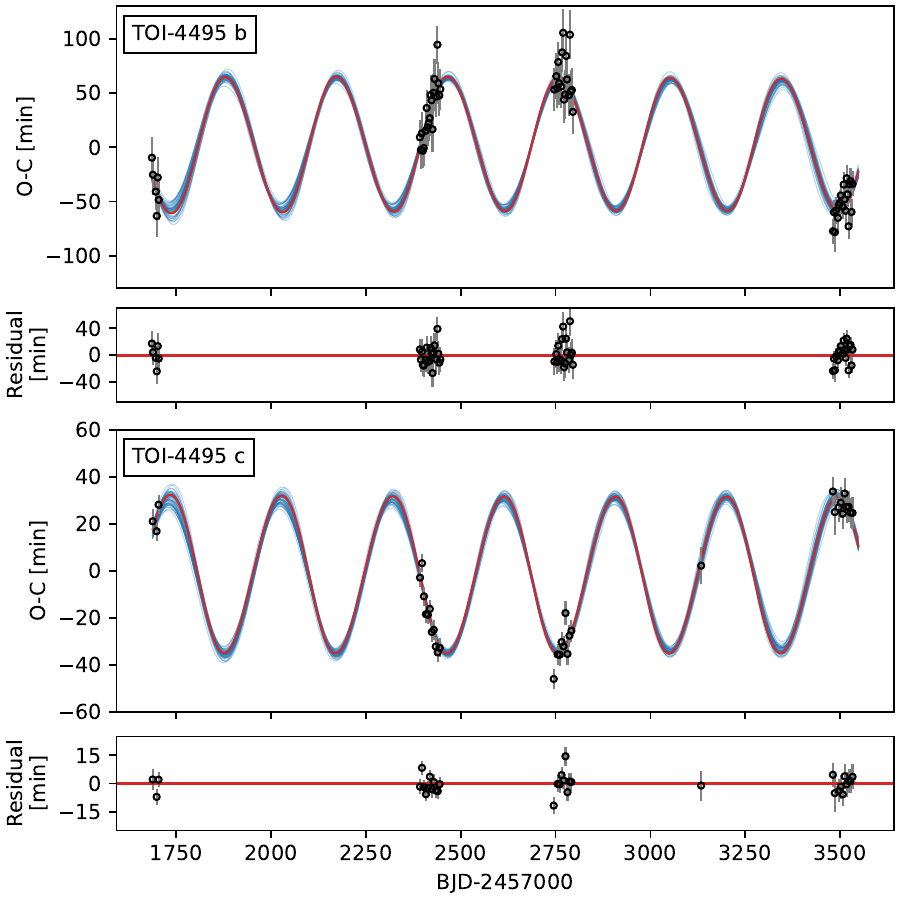}
    \caption{Transit timing variations of TOI-4495 b and c. All observations are from TESS, except for the transit epoch of planet c near day 3100, which is from the KPF RM observation. The blue curves show TTV models from 50 random samples drawn from the photodynamical posteriors (Table \ref{tab:fitted_and_derived_physical_parameter}), the red curve shows the best-fit solution, and the residuals are observed TTV minus the best-fit solution.}
    \label{fig:ttv}
\end{figure*}

\begin{deluxetable*}{lllll}
\tablewidth{1.0\textwidth}
\tablecaption{Fitted and derived physical parameters for TOI-4495 planets. Reported values are median with one $\sigma$ uncertainties; upper limits are quoted at the 95\% credible level. Osculating parameters are referenced to BJD = 2458685.0. \label{tab:fitted_and_derived_physical_parameter}}
\tablehead{\colhead{Parameter} & \colhead{Prior} & \colhead{Posterior}  } 
\startdata
\textit{Stellar Parameter} & & & &\\    
$M_*$ [M$_\odot$]       &   $\mathcal{N}(1.247,0.045)$         &   $1.254^{+0.045}_{-0.045}$          &		&  \\ 
$R_*$ [R$_\odot$]       &   $\mathcal{N}(1.309,0.059)$         &   $1.28^{+0.056}_{-0.04}$            &		&  \\ 
$q_{1}$                 &   $\mathcal{U}(0,1)        $         &   $0.43^{+0.19}_{-0.14}$             &		&  \\ 
$q_{2}$                 &   $\mathcal{U}(0,1)        $         &   $0.17^{+0.23}_{-0.12}$             &		&  \\ 
\hline
\textit{Noise Parameter} & & & &\\ 
$\log a$                &   $\mathcal{U}(-14,-4)$              &   $-12.5^{+1.2}_{-1.0}$                \\ 
$\log c$                &   $\mathcal{U}(-5,1)$                &   $-2.5^{+2.4}_{-1.7}$                 \\ 
$\log \sigma_{\rm TESS}$&   $\mathcal{U}(-4,-4)$               &   $-7.3068^{+0.0044}_{-0.0046}$        \\
$\mu$                   &   $\mathcal{U}(-5\times10^{-4},5\times10^{-4})$ &   $(-2\pm53)\times10^{-7}$  \\
\hline
\textit{Planet Parameter}& & \textit{Planet b} & & \textit{Planet c}\\
$M_p/M_*(\times10^{-5})$ &   $\mathcal{U}( 10^{-2},10^{2} ) $   &   $1.86^{+0.33}_{-0.37}$            &   $\mathcal{U}( 10^{-2},10^{2} ) $   &   $5.6^{+1.0}_{-1.1}$                    \\ 
$R_p/R_*$                &   $\mathcal{U}(0,0.1)            $   &   $0.01782^{+0.00034}_{-0.00034}$   &   $\mathcal{U}(0,0.1)            $   &   $0.02892^{+0.00046}_{-0.00042}$        \\ 
$P_{orb}$ [days]         &   $\mathcal{U}(2.543,2.595)      $   &   $2.56699^{+0.00019}_{-0.0002}$    &   $\mathcal{U}(5.131,5.235)      $   &   $5.18553^{+0.0002}_{-0.00018}$         \\ 
$e_p\cos\omega_p$        &   $\mathcal{U}(-0.2,0.2)         $   &   $-0.075^{+0.011}_{-0.015}$        &   $\mathcal{U}(-0.2,0.2)         $   &   $-0.008^{+0.014}_{-0.044}$             \\ 
$e_p\sin\omega_p$        &   $\mathcal{U}(-0.2,0.2)         $   &   $-0.019^{+0.012}_{-0.023}$        &   $\mathcal{U}(-0.2,0.2)         $   &   $-0.001^{+0.017}_{-0.032}$             \\ 
$b_p$                    &   $\mathcal{U}(0,1)              $   &   $0.26^{+0.13}_{-0.15}$            &   $\mathcal{U}(0,1)              $   &   $0.595^{+0.056}_{-0.046}$              \\ 
$T_{\rm conj}$ [BJD-2457000]&   $\mathcal{U}(1685.73,1685.83)$  &   $1685.7738^{+0.0017}_{-0.0017}$   &   $\mathcal{U}(1687.79,1687.89)  $   &   $1687.8395^{+0.0012}_{-0.0011}$        \\ 
$\Omega_p$ [rad]         &   $\mathcal{U}(-1,1)$                &   $-0.015^{+0.07}_{-0.068}$         &   fixed     					 	 &   $\equiv 0$        						\\ 
\hline
\textit{Derived Parameter} & & & &\\ 
Planetary Mass (M$_\oplus$)		&								& 	 $7.7^{+1.4}_{-1.5}$  			  &									 & 	 $23.2^{+4.4}_{-4.7}$  				    \\ 
Planetary Radius (R$_\oplus$)	&								& 	 $2.483^{+0.138}_{-0.095}$  	  &									 & 	 $4.03^{+0.23}_{-0.15}$  			    \\ 
Bulk Density (g/cm$^3$)			&								& 	 $2.71^{+0.66}_{-0.62}$  		  &									 & 	 $1.90^{+0.47}_{-0.44}$  			    \\ 
Semi-major Axis (AU)			&								& 	 $0.03957^{+0.00047}_{-0.00047}$  &									 & 	 $0.06323^{+0.00075}_{-0.00076}$  	    \\ 
Scaled Semi-major Axis ($a/R_*$)&								& 	 $6.65^{+0.21}_{-0.29}$  		  &									 & 	 $10.63^{+0.34}_{-0.46}$  			    \\ 
Eccentricity					&								& 	 $0.078^{+0.02}_{-0.013}$  		  &									 & 	 $0.023^{+0.043}_{-0.016} (<0.098)$  	\\ 
Inclination (deg)				&								& 	 $87.8^{+1.4}_{-1.3}$  			  &									 & 	 $86.8^{+0.33}_{-0.47}$  			    \\ 
Equilibrium Temperature (K)     &								&    $1735\pm60$                      &									 &   $1365\pm33$                            \\
Mutual Inclination (deg)		&								&	 $3.1^{+3.0}_{-1.7} (<8.7)$ 	  &									 &				--                          \\
Projected Obliquity (deg)       &                               &           -                         &                                  &   $-2.3^{+8.3}_{-7.8}$                   \\
True Obliquity (deg)            &                               &           -                         &                                  &   $<39$                                  \\
\enddata
\end{deluxetable*}

\subsection{Mass, Radius, and Density \label{sec:composition}}

\begin{figure}
\center
\includegraphics[width = 1.\columnwidth]{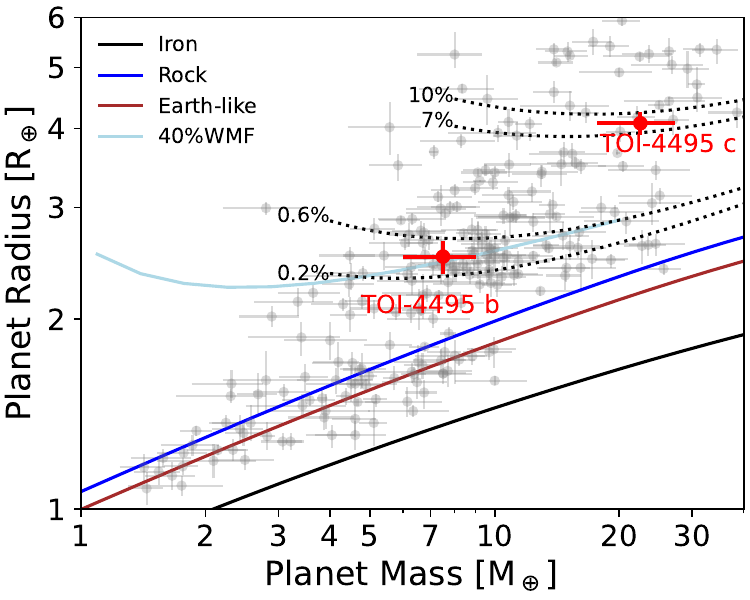}
\caption{Masses and radii of the TOI-4495 planets and other planets with measured precision > $3\sigma$ from the NASA Exoplanet Archive \citep{Christiansen2025_EA}. Both axes are shown in log-scale. Theoretical mass–radius relationships for pure-iron, pure-rock, and Earth-like compositions are shown in black, blue, and brown, respectively \citep{Zeng2016}. TOI-4495 b is consistent with either possessing a thin H/He envelope comprising 0.2–0.6\% of its total mass (black dotted lines) or being an ocean world with a water mass fraction of approximately 40\% (light blue; \citealt{Aguichine}). TOI-4495 c is consistent with hosting a more substantial H/He envelope, accounting for 7–10\% of its total mass.}
\label{fig:mass_radius}
\end{figure}

We show the measured masses and radii of TOI-4495 planets on planetary mass-radius diagram, along with other exoplanets with mass and radius measurement precision better than 25\% from Exoplanet Archive \citep{Christiansen2025_EA}\footnote{Accessed on 2025-5-15} in Figure \ref{fig:mass_radius}. TOI-4495 b is a sub-Neptune with a radius of $2.48\pm0.14$ R$_\oplus$ and a mass of $7.7\pm1.4$ M$_\oplus$, corresponding to a bulk density of $2.71\pm0.64\,{\rm g\,cm^{-3}}$. TOI-4495 c is a Neptune-like planet with a radius of $4.03\pm0.19$ R$_\oplus$ and a mass of $23.2\pm4.5$ M$_\oplus$, yielding a mean density of $1.90\pm0.45\,{\rm g\,cm^{-3}}$. The low densities of both planets suggest the presence of volatile-rich atmospheres.

Taking their insolation into account, we applied the parametric model of \cite{Chen_Rogers} to estimate their envelope masses, assuming pure hydrogen–helium compositions. TOI-4495 b is consistent with a thin H/He envelope comprising 0.2–0.6\% of its mass, while TOI-4495 c appears to host a thicker atmosphere containing 7.5–10\% of its total mass. We note, however, that the \cite{Chen_Rogers} model does not include internal energy sources such as tidal heating. As discussed in Section~\ref{sec:tide}, TOI-4495 b may experience significant tidal heating, which could contribute to its inflated radius. In that case, the inferred envelope mass fraction for TOI-4495 b should be regarded as an upper limit. Alternatively, TOI-4495 b could be consistent with a water-rich atmosphere, as suggested by interior structure models that allow for high-metallicity envelopes \citep[e.g.,][]{Aguichine}.

\subsection{Dynamics and Mean Motion Resonance}

In Figure \ref{fig:ttv}, we show the predicted transit timing variations derived from the posteriors of our photodynamical model. Using the average orbital periods obtained by forward-modeling the posterior samples, we calculated the normalized distance to resonance to be $\Delta = 0.008765 \pm 3\times10^{-6}$, corresponding to a super-period of $295.6 \pm 0.1$ days. A sinusoidal fit to the observed TTVs yields a period of $293.3 \pm 0.8$ days, consistent with the predicted super-period and confirming that the observed timing variations are dominated by near-resonant dynamics \citep{Lithwick_ttv}.

\begin{figure*}
    \centering
    \label{fig:best_fit_1000_kyr}
    \includegraphics[width = 0.99\textwidth]{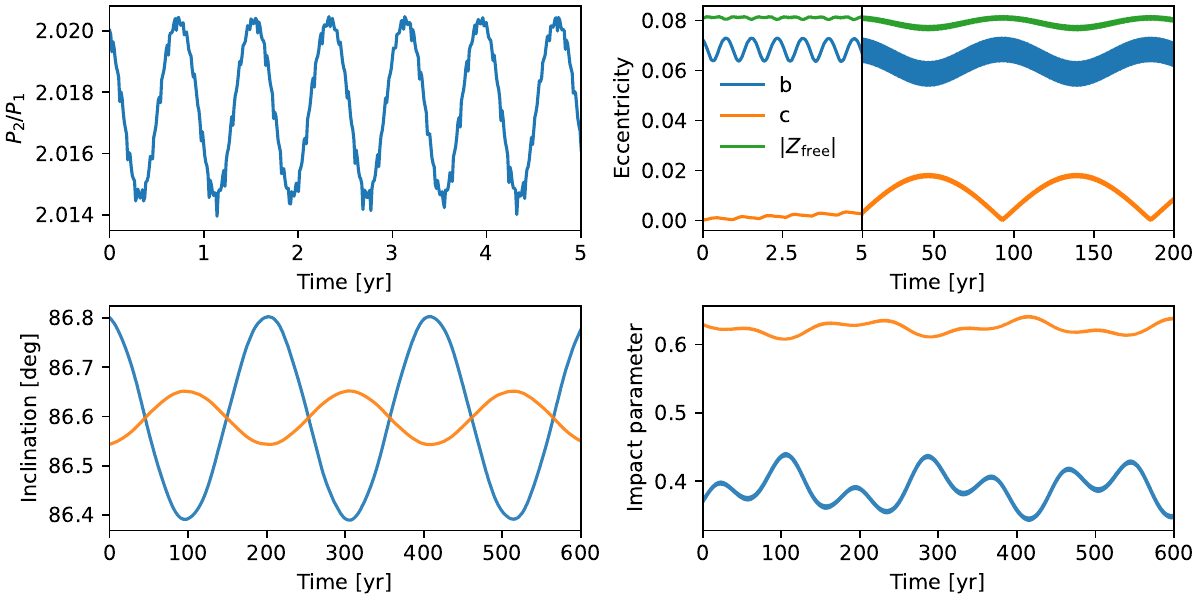}
\caption{Temporal evolution behavior of the period ratios, eccentricity, inclination, and impact parameter of TOI-4495 planets. The integrated orbit is from the best-fit solution. Notice the horizontal axis of the top right panel has two different scales.}
\end{figure*}

We forward-integrated the best-fit solution for 600 yr to examine the short-term dynamical behavior of the TOI-4495 pair, as shown in Figure \ref{fig:best_fit_1000_kyr}. The period ratio of the planets oscillates between a minimum of 2.014 and a maximum of 2.020 on the timescale of the super-period. On a similar timescale, the planets exchange eccentricity with an amplitude forced by resonant interactions (Eq. \ref{eq:forced_ecc_1}, \ref{eq:forced_ecc_2}). A slower modulation due to secular interactions occurs on a timescale of $\sim$90 years. The free eccentricity component $|Z_{\rm free}|$ (Equation \ref{eq:comb_ecc}) remains nearly constant over the observational baseline but varies in orientation and magnitude on the secular timescale.

Direct modeling of the light curves allows us to leverage transit duration variations (or their absence) to constrain the mutual inclination $\Delta I$ between the planetary orbits, calculated as
\begin{equation}
    \cos \Delta I = \cos i_b \cos i_c + \sin i_b\sin i_c\cos(\Omega_c-\Omega_b).
\end{equation}
The posteriors suggest a low mutual inclination of $\Delta I=3.1^{+3.0}_{-1.7}$\degree, with an upper limit of $8.7\degree$ at the 95\% credible level. \footnote{However, since we imposed positive prior on impact parameters for both planets, the mutual inclination obtained here is likely underestimated.} The best-fit solution yields $\Delta I=0.28\degree$. In the lower-left panel of Figure \ref{fig:best_fit_1000_kyr}, we find that the orbital inclinations of both planets oscillate on a timescale of $\sim$200 yr, consistent with the nodal precession timescale of 211 yr predicted by Laplace–Lagrange secular theory \citep[][Equation~7.31]{MurrayDermott}. Because the mutual inclination is very small, the amplitude of these oscillations is negligible, and the impact parameters of both planets remain below unity, ensuring a permanent transit configuration.

The combined eccentricity $|Z|$ (Equation \ref{eq:comb_ecc}) from the photodynamical posteriors is $0.085^{+0.02}_{-0.01}$, well below the critical value of $|Z|\sim0.23$ required for the onset of chaos due to MMR overlap \citep[][Eq.~19]{Hadden2018}. We randomly selected 109 posterior samples and integrated them for 1 Myr using the \texttt{TRACE} integrator \citep{TRACE} within \texttt{rebound}, finding no instances of instability.

\begin{figure*}
    \hspace{-0.2cm}
    \label{fig:hamiltonian}
    \includegraphics[width = 0.44\textwidth]{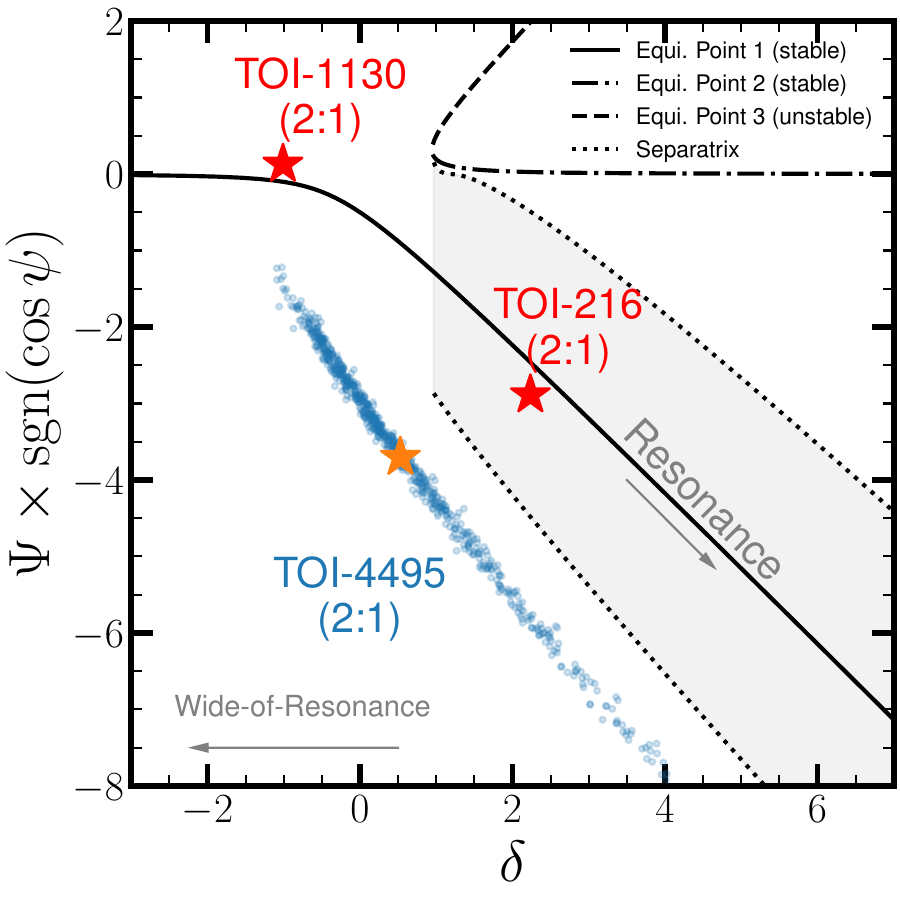}
    \hspace{1.2cm}
    \includegraphics[width = 0.455\textwidth]{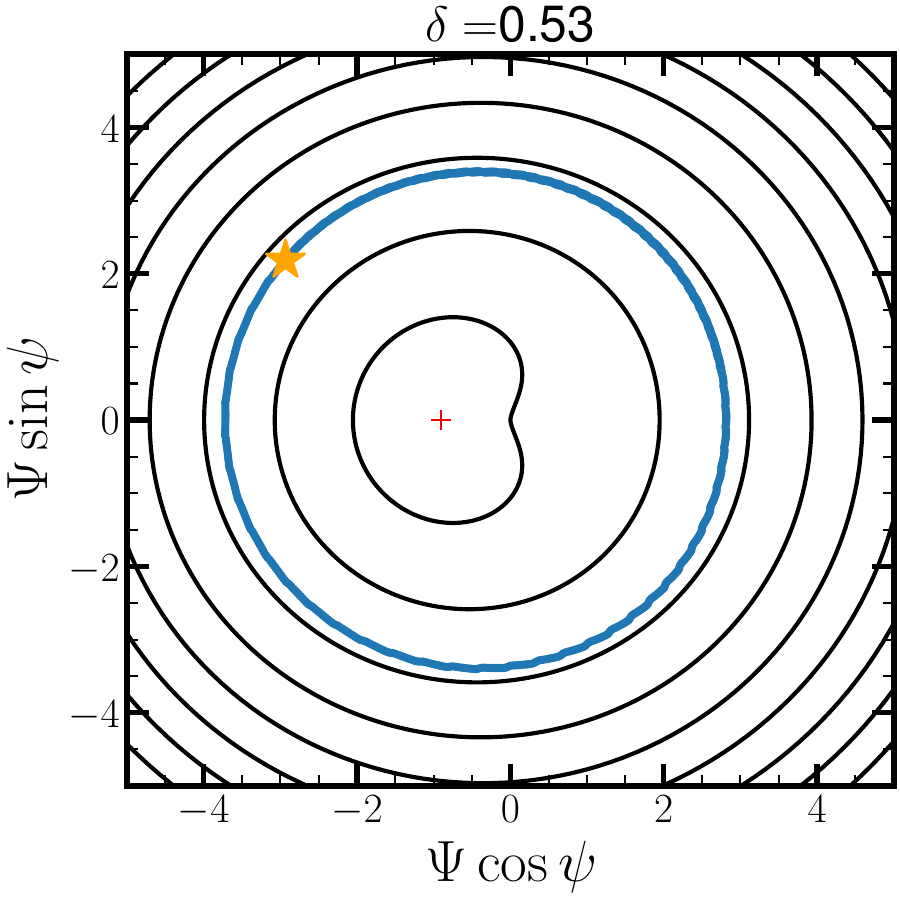}
\caption{Hamiltonian of first-order mean motion resonance, adapted from \cite{Nesvorny}. {\it Left}: The location of fixed points (solid, dashed, and dotted-dashed lines) and separatrix (dotted lines) as a function of $\delta$, a parameter measures the proximity to resonance. $\psi$ is the resonant angle, and its conjugate momentum, $\Psi$, scales with combined eccentricity of planets. Resonant orbits, shown as shaded area and enclosed by the separatrix, exist in $\delta>0.95$. Blue points are the posterior of TOI-4495 from the photodynamics analysis, which has $\delta=0.4^{+1.9}_{-0.7}$. Orange star is the best-fit solution. We also show the resonant TOI-216 and near-resonant TOI-1130 for comparison. {\it Right}: the phase space at $\delta=0.53$, the blue trajectory is the integrated orbit of best-fit solution (orange star in left figure) which suggest a circulating solution. Red cross is the stable equilibrium point (solid line on the left).}
\end{figure*}

The planets in the TOI-4495 system are close to the 2:1 MMR. To assess whether TOI-4495 is in resonance, we mapped its orbital configuration onto the resonant Hamiltonian framework. The first-order $j:(j-1)$ resonance shares the same Hamiltonian \citep[e.g.,][]{Nesvorny}:
\begin{equation}
    H = -(\Psi - \delta)^2 - \sqrt{2\Psi} \cos \psi,
\end{equation}
where the action $\Psi$ scales with the combined eccentricity $Z$ (Equation \ref{eq:comb_ecc}), and its conjugate angle $\psi=j\lambda_2-(j-1)\lambda_1 -\hat{\varpi}$ is the resonant angle, with the mixed pericenter angle $\hat{\varpi}=\arg(Z)$ \citep{Petit,Dai2023}. The conserved quantity $\delta$ represents the system’s proximity to resonance and determines the topology of the Hamiltonian phase space, as shown in the left panel of Figure \ref{fig:hamiltonian}. The $\delta$ parameter is determined by planet mass, eccentricity and period ratios, and should not be confused with the $\Delta=(j-1)P_2/(jP_1)-1$, which depend solely on period ratios. Formally resonant orbits, depicted as the gray region enclosed by separatrices, exist only after the bifurcation of fixed points, which emerge when $\delta>0.945$. These resonant orbits exhibit librating resonant angles. For $\delta<0.945$, orbits with librating angles still exist but remain near the stable fixed points. 

Our photodynamical posteriors yield a proximity parameter of $\delta=0.4^{+1.9}_{-0.7}$, spanning values both before and after the bifurcation. However, the posterior samples correspond to sufficiently large action (free eccentricity) that they lie in the external circulation zone. The trajectory of the best-fit model, shown in the right panel of Figure~\ref{fig:hamiltonian}, confirms that the resonant angle $\psi$ circulates over time. In contrast to systems such as TOI-1130 \citep{Korth2023} and TOI-216 \citep{Dawson2021}, which reside near resonant equilibrium points in their phase space, TOI-4495 lies far from resonance. This indicates that its free eccentricity component dominates over the forced component, suggesting past dynamical excitation and a departure from resonance capture. We discuss possible origins in Section \ref{sec:overstable}.

\subsection{Eccentricity and Tides \label{sec:tide}}

\begin{figure}
    \hspace{-0.1cm}
    \includegraphics[width=0.95\linewidth]{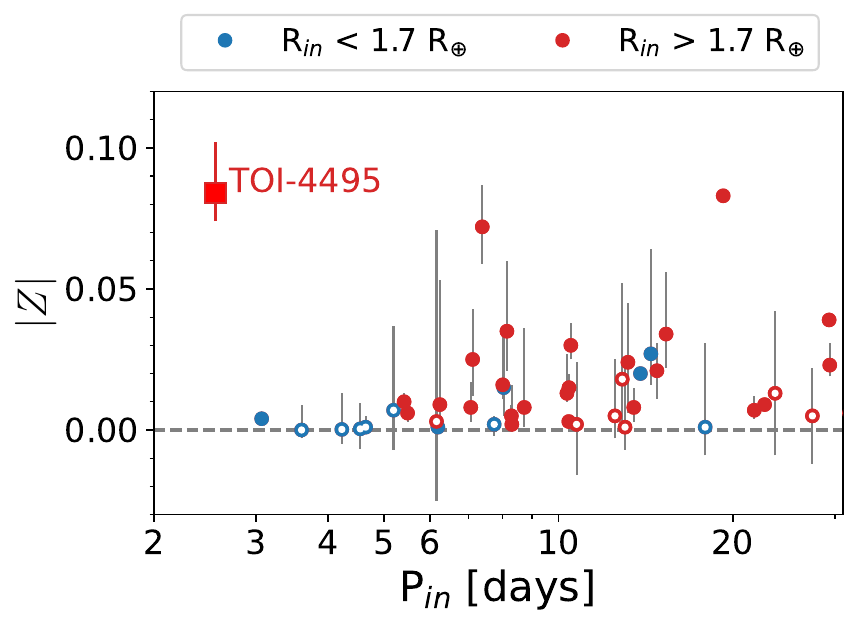}
    \caption{Eccentricities, $|Z|$, as a function of the innermost period $P_{\rm in}$ of TOI-4495 (square) and 43 pairs of ${\it Kepler}$ near-resonant planets (circles) from \cite{Hadden2017}. Anticipating different tidal dissipation rates on terrestrial ($R_p<1.7~R_{\oplus}$) and gaseous ($R_p>1.7~R_{\oplus}$) planets, we mark them in red and blue color, respectively. $|Z|$ consistent with zero are shown as open circles.}
    \label{fig:hadden17_ecc_period}
\end{figure}

Our dynamical model suggest TOI-4495 b possess non-zero eccentricity $e_b=0.078^{+0.02}_{-0.013}$, while planet c has an upper limit of $e_c<0.098$ at 95\% confidence interval. This eccentricity is significant and consistent with the upper end of the population-level eccentricity distribution of multi-planet systems \citep{Lithwick_ttv,Xie,vaneylen2015}.  For near-resonant planets, the free eccentricity is commonly non-zero and fall in the range of 0.01-0.04 \citep{Hadden2013,Hadden2017}. However, TOI-4495 is a close-in ($a/R_*\lesssim10$) system, therefore tidal forces may be important in shaping the orbital characteristics of the system, which is the focus of discussion for this section.

In Figure \ref{fig:hadden17_ecc_period} we compare the eccentricity\footnote{Here we compare the projected quantity $Z_{\rm proj} = ZZ^*_{\rm med}/|Z|$ as defined in \cite{Hadden2017}, where $Z_{\rm med}$ is the median of the real and imaginary components of $Z$ from the posterior. This regularization highlights systems with eccentricities consistent with zero, and thus the lower limit can be negative.} and orbital period of the TOI-4495 system with near-resonant planet pairs in the \emph{Kepler} sample \citep{Hadden2017}. 
For the \emph{Kepler} sample, planetary eccentricities are largely consistent with zero when the orbital period of the innermost planet is shorter than 5 days. A closer inspection of the radii of the inner planets in these pairs reveals that they are typically smaller than 1.7 $R_\oplus$ and are therefore likely rocky \citep{Weiss2014,Rogers,Fulton,Dressing}. TOI-4495 is thus a rare near-resonant system with large free eccentricity at such short orbital periods. This contrast can be attributed to differences in tidal dissipation rates between terrestrial and gaseous planets. The damping timescale due to equilibrium eccentricity tides raised on a planet by its host star is \citep{Yoder1981}:

\begin{align}
\label{eq:tidal_damping_timescale }
    \tau_e \equiv \frac{e}{\dot{e}}&\approx0.7 {\rm~Myr}  
    \left(\frac{Q'}{10^3}\right)
    \left(\frac{M_p/M_*}{M_\oplus/M_\odot}\right)     \\ \nonumber
    &\left(\frac{R_*/R_p}{R_\odot/R_\oplus}\right)^5
    \left(\frac{\rho_*}{\rho_\odot}\right)^{5/3}
    \left(\frac{P_{\rm orb}}{\rm 1~day}\right)^{13/3} \\
    &\approx 14~{\rm Myr}~(\frac{Q'}{10^3}) {\rm~for~planet~b}
\end{align}
where $Q'=3Q/2k_2$ is the reduced tidal quality factor, which reflects the body's ability to dissipate tidal energy and is determined by the internal structure, rheology, and thermal state of the planet. In general, $Q'$ is anti-correlated with the core radius and shear modulus \citep{Lainey2016}, and thus terrestrial planets are more efficient in energy damping and have shorter eccentricity damping timescale. In Solar system, terrestrial planets are 2-3 orders of magnitude quicker in dissipation compared to simplified expectations for gaseous planet \citep[$Q'_{\rm Mars}\sim100-1000,~Q'_{\rm Jupiter}\sim10^5$;][]{Goldreich1966_Q}. Recently, \cite{Louden2023ApJ} investigated the dissipation rate for the super-earth and sub-Neptunes near MMRs and find two tentative peaks at $Q'=10^2$ and $10^6$, suggesting that distinct dissipation regimes also exist for these geophysically different populations. 

If TOI-4495 b is indeed still in the course of circularizing eccentricity gained at formation, i.e., circularization timescale is similar to or greater than system age $\tau_e\gtrsim1.9$ Gyr, this would imply $Q'\gtrsim10^5$, which is comparable to the $Q'$ of gaseous planets in Solar system. Adopting this fiducial value, we estimate the present-day tidal heating rate following the Equation 7 in \cite{Levrard2007_tide}, assuming zero planet obliquity. This returns an upper limit on the tidal heating rate of $3\times10^{17}$ W. The tidal heating per unit mass ($6\times10^{-9}$ W/kg) is comparable that of Io ($\sim10^{-10}-10^{-9}$ W/kg, \cite{Veeder_Io}). The tidal luminosity is $10^{-3}$ of the irradiation received by TOI-4495 b ($L_{\rm tide}=10^{-3}L_{\rm ins}$). \citet{Millholland2019} suggest that sub-Neptune planets receiving tidal heating at a level exceeding $10^{-5}L_{\rm ins}$ may undergo inflation of their envelopes. Using the parametric model of \citet{Millholland2019} (their Equation 14), we find that even at an insolation level of 500 Earth fluxes (500 $F_\oplus$), a planet with an envelope mass fraction $f_{\rm env}$ of only 0.3\% would be inflated to 1.5 times the size predicted without tidal heating. This factor matches the observed radius-to-core radius ratio for TOI-4495 b (2.5/1.7). Because the inflation effect strengthens with higher insolation, and TOI-4495 b receives 1450 $F_\oplus$, we conservatively adopt an upper limit of $f_{\rm env}<0.3\%$ for a pure H/He envelope.

\subsection{Formation Constraints from Disk Migration \label{sec:overstable}}

TOI-4495 hosts a pair of near-resonant sub-Neptunes with coplanar orbits aligned with the stellar rotation axis. This orbital architecture and near-resonant configuration suggest a formation scenario in which both planets underwent convergent migration within the protoplanetary disk. However, a key discrepancy in this picture is the moderate eccentricity observed for planet b. Migration simulations \citep{Keller2025} suggest that planets in resonant chains can attain eccentricities up to 0.1, but these eccentricities are primarily excited by resonant interactions (forced eccentricity), unlike TOI-4495 where the eccentricities are dominated by the free component. Such eccentricities are expected to be damped during the disk phase via planet–disk interactions. This indicates that additional mechanisms are required to excite or maintain the eccentricity of planet b.

Can the observed eccentricity be primordially excited by resonant capture? Planets can indeed attain a range of equilibrium eccentricities during resonant capture \citep[e.g.,][]{Goldberg2023}. However, beyond a certain limit, resonant capture becomes unstable and the planet may escape from the MMR. The stability of resonant capture depends on planet properties and the local dissipation rate in a complex way, but it generally weakens at low-order MMRs (e.g., 2:1), for low-mass planets with small inner-to-outer mass ratios ($q$), and when eccentricity damping is weak relative to migration (low $K=\tau_a/\tau_e$). These conditions are highly relevant for TOI-4495, which lies near the 2:1 MMR and has $q\sim0.3$.

\begin{figure*}
    \centering
    \includegraphics[width=0.9\textwidth]{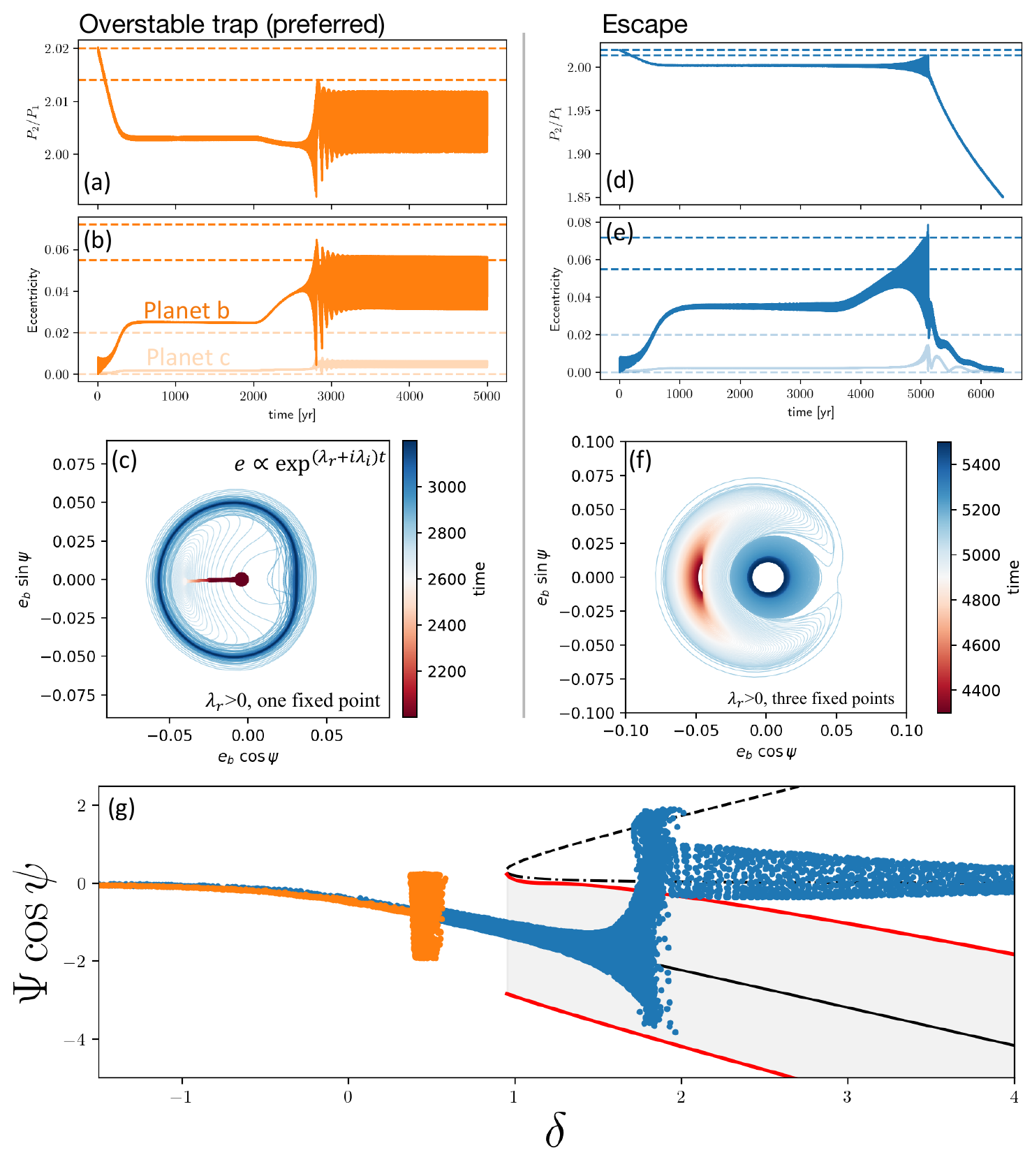}
    \caption{Outcomes of the dissipative resonant evolution of TOI-4495 analogues, categorized by the stability of equilibrium points and the number of fixed points at capture. Panels (a–c): Period ratio, eccentricity, and resonant angle $\psi$ during an overstable trap. In this regime, the real part of the eigenvalue is positive ($\lambda_r>0$) but the resonant fixed point has not bifurcated. Eccentricities are excited and saturate at finite values, leading to wide oscillations around the equilibrium. Panels (d–f): The same quantities for resonant escape. Here, the real part of the eigenvalue is positive and the conservative fixed point has bifurcated. Eccentricities continue to grow until the system crosses into the inner circulation region, moving narrowly away from the 2:1 MMR. The planets then escape resonance and migrate toward more compact commensurabilities. The dashed lines in the first two columns mark the observed period ratio and eccentricities of TOI-4495. Panel (g): Phase-space evolution of the two cases. Both satisfy the analytic overstable condition ($\lambda_r>0$). In the trap case, the system reaches equilibrium before bifurcation ($\delta < 0.95$), while in the escape case the trajectory evolves beyond bifurcation ($\delta > 0.95$), resulting in resonant departure.} 
    \label{fig:trap}
\end{figure*}

Linear stability analysis around the equilibrium point of resonant capture shows that perturbations to the eccentricity evolve as $e\propto \exp(\lambda t)$, where $\lambda=\lambda_r+i\lambda_i$ is the characteristic growth frequency \citep{Goldreich2014,Deck2015_overstable,Xu0217,lin2025_overstable}. If $\lambda_r<0$, eccentricity behaves as a damped oscillator and settles to a fixed point with small libration amplitude. If $\lambda_r>0$, eccentricity is further excited, following an overstable out-spiral (Figure \ref{fig:trap}-c). The $\lambda_r>0$ occurs for $K<K_{\rm overstable}$ \citep{lin2025_overstable}:
\begin{equation}
    K_{\rm overstable} = \left[\frac{3(j-1)}{\mu_2f\alpha}\right]^{2/3} f(q)h(q) \sim 710,
\end{equation}
where $\mu_2$ is the outer-to-stellar mass ratio, $\alpha=a_1/a_2$ is the semi-major axis ratio of the planets, and $f(q), h(q)$ are functions of the mass ratio $q$ \citep[][Eq. 30, 31]{lin2025_overstable}. The overstable trap can persist indefinitely (Figure \ref{fig:trap}-a), provided the fixed points in the conservative problem have not bifurcated ($\delta<0.95$; or the orange trajectory in Figure \ref{fig:trap}-g). Once bifurcation occurs, a new fixed point emerges in the inner circulation region (corresponding to $e\sim0$). In that case, the unstable out-spiral no longer saturates but slides into the circulation region (Figure \ref{fig:trap}-f). Because the resonant interaction weakens as eccentricity decreases, the planet can no longer be stalled at the MMR, leading to resonance escape (Figure \ref{fig:trap}-d, and the blue trajectory in panel-g). The bifurcation threshold is given by
\begin{equation}
\label{eq:escape}
    K_{\rm esc} = \left[\frac{3[(j-1)^2+j^2q]}{\mu_2(f\alpha+gq^2)}\right]^{2/3} \frac{h(q)}{4} \sim 292,
\end{equation}
and resonance escape requires the system to already be in the overstable regime, which is always true for TOI-4495 since $K_{\rm esc}<K_{\rm overstable}$.

Based on these trapping criteria, we can estimate the maximum eccentricity the inner planet can acquire during resonant capture. The equilibrium eccentricity of the inner planet is related to $K$ by \citep{lin2025_overstable}:
\begin{equation}
    e_{\rm 1,eq} = \left(\frac{1}{1+q\sqrt{\alpha}} \frac{f^2}{jf^2+(j-1)g^2q\sqrt{\alpha}} K^{-1}\right)^{1/2}.
\end{equation}
Substituting $K_{\rm esc}$ into this expression yields $e_{\rm i,max}\sim0.04$, consistent with numerical verification. Thus, if the observed eccentricity of planet b were solely due to resonant capture, disk dispersal would have had to occur during the onset of instability, on a timescale of $\sim0.1\tau_a$ \citep{Deck2015_overstable}—much shorter than typical disk lifetimes, posing a fine-tuning problem.

In the Type-I migration regime, the relative strengths of migration and damping depend on disk properties \citep{Tanaka2004,Kley_2012}. The MMR trapping criteria therefore provide constraints on disk conditions that could form TOI-4495 analogues \citep[e.g.,][]{Delisle2015_overstable,Huang_capture}. To explore these constraints, we carried out a suite of convergent migration simulations using \texttt{rebound} with the \texttt{type\_I\_migration} module in \texttt{reboundx}. We adopted a static disk profile $\Sigma=\Sigma_{\rm 1AU}(r/{\rm AU})^{-s}$ and $h=h_{\rm 1AU}(r/{\rm AU})^\beta$ with $s=1$ and $\beta=0.25$. Planets with masses from the TTV measurements were initialized slightly wide of the 2:1 MMR. 

\begin{figure}
    \centering
    \includegraphics[width=1.0\columnwidth]{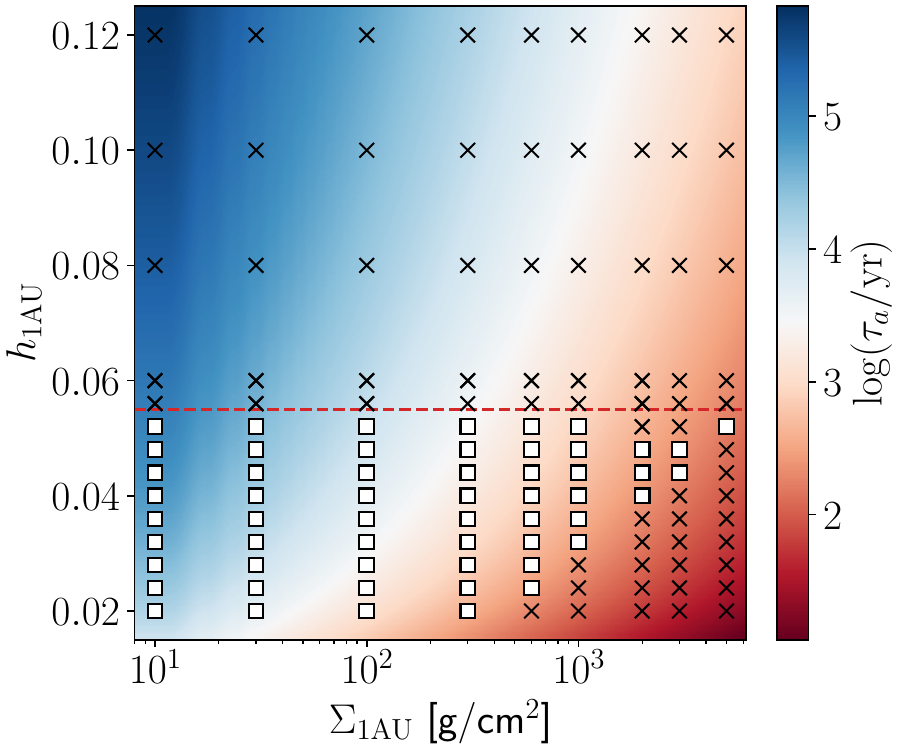}
    \caption{The surface density and aspect ratio at 1 AU of the disk that produce 2:1 MMR trappings of TOI-4495 analogues. Squares denote 2:1 capture (the observed architecture) while crosses denotes escape. The background color denotes the migration rate at the observed location of TOI-4495 ($\sim 0.06$ au). The red dashed line shows the critical height aspect ratio above which the capture will become subject to escape (Eq. \ref{eq:escape} and \ref{eq:scale_height}).}
    \label{fig:sf_h}
\end{figure}

The ratio $K=\tau_a/\tau_e$ is primarily determined by the local disk aspect ratio $h$ \citep{Kajtazi2023}:
\begin{equation}
\label{eq:scale_height}
    h = \left(\frac{0.78}{2.7+1.1s}K^{-1}\right)^{1/2}.
\end{equation}
Substituting $K_{\rm esc}$ gives a maximum aspect ratio of $h_{\rm max}\sim0.026$ at 0.06 au, corresponding to $h_{\rm 1au}\sim0.055$ at 1 au. We performed a grid of Type-I migration simulations with $\Sigma_{\rm 1AU}\in[10,5000]$ g\,cm$^{-2}$ and $h_{\rm 1AU}\in[0.02,0.12]$, integrating each run for at least one $\tau_a$. The outcomes, summarized in Figure~\ref{fig:sf_h}, show that trapping near the 2:1 MMR occurs only in low-density, thin disks with $\Sigma_{\rm 1AU}<5000$ g\,cm$^{-2}$ and $h_{\rm 1AU}\lesssim0.058$. The upper boundary of successful capture on $h$ is well described by criterion of overstability, outlined by the horizontal red dashed lines, while the lower boundary of $h$ is dictated by the adiabatic theory and/or capture stability under strongly dissipative conditions \citep{Batygin2015,Huang_capture,Batygin2023}.


Despite successful resonance capture under favorable disk conditions, the resulting period ratios and eccentricities in our simulations remain lower than observed (dashed curves in the second row of Figure~\ref{fig:trap}). This indicates that additional processes are needed to further separate the orbits and excite the eccentricities. Possible contributors include stochastic forcing from turbulent density fluctuations in the disk \citep{Adams2008,Petigura2018,Goldberg2023,Chen2025_turbulent} and post-disk planetesimal impacts on the inner planet \citep{Wu2024,Li2025}. Alternatively, the fact that planet c is significantly more massive and has a smaller eccentricity may indicate angular momentum deficit equipartition, possibly involving an undetected outer companion \citep{Choksi_TTVphase}.

\section{Conclusion\label{sec:concl}}
In this paper, we report the confirmation and follow-up observation of TOI-4495, a pair of near-resonant sub-Neptunes aligned with their host stars. The main conclusions of the paper are summarized below:
\begin{itemize}
    \item TOI-4495 b is a sub-Neptune with a mass of $M_b=7.7\pm1.4$ M$_\oplus$ and a radius of $R_b=2.48^{+0.14}_{-0.10}$ R$_\oplus$, while TOI-4495 c is a Neptune-like planet with $M_c=23.2\pm4.5$ M$_\oplus$ and $R_c=4.03^{+0.23}_{-0.15}$ R$_\oplus$ (Figure \ref{fig:mass_radius}). Both planets likely host volatile-rich atmospheres, either solar-metallicity H/He-dominated envelopes or water-enriched compositions.
    
    \item Spectroscopic follow-up of planet c revealed a well-aligned orbit with the stellar rotation axis, with a sky-projected obliquity of $-2.3^{+8.3}_{-7.8}~^\circ$ (Figure \ref{fig:RM_DT}). Our photodynamical model constrained the mutual inclination between planets b and c to be $<8.7^\circ$ (95\% credible upper limit). Thus, the system is likely coplanar and aligned, adding to the growing sample of well-aligned multi-transiting systems \citep{Zhang2025}.
    
    \item No significant excess H$\alpha$ absorption was detected during the transit of TOI-4495 c, suggesting a low or suppressed atmospheric mass-loss rate, consistent with its high mass and the host’s low stellar activity.
    
    \item The planets lie near the 2:1 MMR ($\Delta=0.008765\pm3\times10^{-6}$) and exhibit transit timing variations (Figure \ref{fig:ttv}), but they are not resonant: their resonant arguments circulate (Figure \ref{fig:hamiltonian}).
    
    \item TOI-4495 b has a significant eccentricity of $e_b=0.078^{+0.020}_{-0.013}$, supported by its large TTV phase (Figure \ref{fig:ttv_phaseshift}). Interpreting this as ongoing tidal circularization implies a reduced tidal quality factor of $Q'\gtrsim10^5$, consistent with expectations for sub-Neptunes.
    
    \item Dynamical simulations show that overstable resonant capture could stall the system near, but just outside, the 2:1 resonance. However, this mechanism only partially accounts for the observed free eccentricity ($\sim0.04$ compared to the measured $e_b=0.078^{+0.021}_{-0.013}$). Additional eccentricity excitation mechanisms are therefore likely required.
\end{itemize}

\software{{\sc AstroImage} \citep{Collins:2017}, {\sc Isoclassify} \citep{Huber}, {\sc isochrones} \citep{Morton}, {\sc MIST} \citep{MIST}, {\sc SpecMatch-Syn} \citep{Petigura_thesis}, {\sc Batman} \citep{Kreidberg2015}, {\sc emcee} \citep{emcee}}, {\sc jnkepler} \citep{2025ascl.soft05006M}

\facilities{Keck I: (KPF), {\it TESS}}

\begin{acknowledgments}
\begin{center}
ACKNOWLEDGEMENTS
\end{center}

We thank the referee for constructive feedback that improves the quality and presentation of the manuscript. We also thank Konstantin Batygin, Nick Choksi, Beibei Liu, Linghong Lin, Wenrui Xu, and Mike Greklek-McKeon for helpful discussions. This work is supported by National Key R\&D Program of China, No.2024YFA1611801, the National Natural Science Foundation of China (grant No.124B2058), and science research grants from the China Manned Space Project with No. CMSCSST-2025-A16. Support for this work was provided by NASA through the NASA Hubble Fellowship grant HST-HF2-51503.001-A awarded by the Space Telescope Science Institute, which is operated by the Association of Universities for Research in Astronomy, Incorporated, under NASA contract NAS5-26555. Work by K.M. was supported by JSPS KAKENHI grant No.~25K07387. D.R. was supported by NASA under award number 80NSSC25M7110..

This work was supported by a NASA Keck PI Data Award, administered by the NASA Exoplanet Science Institute. Data presented herein were obtained at the W. M. Keck Observatory from telescope time allocated to the National Aeronautics and Space Administration through the agency's scientific partnership with the California Institute of Technology and the University of California. The Observatory was made possible by the generous financial support of the W. M. Keck Foundation. The research was carried out, in part, at the Jet Propulsion Laboratory, California Institute of Technology, under a contract with the National Aeronautics and Space Administration (80NM0018D0004)

The data presented herein were obtained at the W. M. Keck Observatory, which is operated as a scientific partnership among the California Institute of Technology, the University of California and the National Aeronautics and Space Administration. The Observatory was made possible by the generous financial support of the W. M. Keck Foundation.

The authors wish to recognize and acknowledge the very significant cultural role and reverence that the summit of Maunakea has always had within the indigenous Hawaiian community.  We are most fortunate to have the opportunity to conduct observations from this mountain.

We acknowledge the use of public TESS data from pipelines at the TESS Science Office and at the TESS Science Processing Operations Center. Resources supporting this work were provided by the NASA High-End Computing (HEC) Program through the NASA Advanced Supercomputing (NAS) Division at Ames Research Center for the production of the SPOC data products.

This paper made use of data collected by the TESS mission and are publicly available from the Mikulski Archive for Space Telescopes (MAST) operated by the Space Telescope Science Institute (STScI) and can be accessed via \cite{tess_2minute}.

\end{acknowledgments}

\bibliography{main}


\counterwithin{figure}{section}

\end{document}